\documentclass[twocolumn,showpacs,preprintnumbers,amsmath,amssymb]{revtex4}

\usepackage{graphicx}
\usepackage{dcolumn}
\usepackage{bm}

\begin{document}

\preprint{APS/123-QED}

\title{Accurate Monte Carlo critical exponents for Ising lattices}

\author{Jorge Garc\'{\i}a}
 \email{jg.garcia@uam.es}
\author{Julio A. Gonzalo}%
 \email{julio.gonzalo@uam.es}
\affiliation{%
Departamento de F\'\i sica de Materiales, Universidad Aut\'onoma de Madrid\\
Cantoblanco, 28049 Madrid, Spain.
}

\author{Manuel I. Marqu\'es}
 \email{manuel@argento.bu.edu}
\affiliation{
Center for Polymer Studies and Department of Physics, Boston University\\
Boston MA 02215, USA.
}

\date{\today}

\begin{abstract}
A careful Monte Carlo investigation of the phase transition very  
close to the critical point ($T \longrightarrow T_c$, $H \longrightarrow$ 0) in
relatively large d = 3, s = 1/2 Ising lattices did produce critical exponents
$\beta_{3D}$ = 0.3126(4) $\cong$ 5/16, $\delta_{3D}^{-1}$ = 0.1997(4) $\cong$
1/5 and $\gamma_{3D}$ = 1.253(4) $\cong$ 5/4. Our results indicate that, 
within experimental
error, they are given by simple fractions corresponding to the linear
interpolations between the respective two-dimensional (Onsager) and
four-dimensional (mean field) critical exponents. An analysis of our inverse
susceptibility data $\chi^{-1}(T)$ vs. $\mid$$T - T_c$$\mid$ shows that these
data lead to a value of $\gamma_{3D}$ compatible with $\gamma'= \gamma$ and
$T_c$ = 4.51152(12), while $\gamma$ values obtained recently by high and low
temperature series expansions and renormalization group methods are not.

\end{abstract}

\maketitle

\section{\label{sec:level1}Introduction\protect}

The Ising model\cite{Adler}, rightly considered as the prototype of statistical
systems  with non-classic power law critical behavior, has been extensively
investigated for many years. Systems with short-range interactions display
Ising-like critical behavior, f.i. liquid-vapor, multicomponent fluid mixtures,
uniaxial magnets, etc, and there is a wealth of very accurate experimental
information on these systems.

To describe f.i. the behavior of a $s = 1/2$
uniaxial ferromagnet near the critical point ($T = T_c$, $H = 0$) two
critical exponents, $\beta$ for the spontaneous magnetization $M_{s}(T)$ and 
$\delta^{-1}$ for the field dependence of the magnetization $M_{c}(H)$ at the
critical temperature, determine basically the critical behavior through
$M_{s}(T) \sim$ $\mid$$T_{c} - T$$\mid^{\beta}$ and $M_{c}(H) \sim H^{1/\delta}$.
It is well known\cite{Onsager} that for a two dimensional Ising lattice ($d$ =
2) Onsager's solution gives fractional values for $\beta_{2D} = 1/8$ and
$\delta_{2D}^{-1} = 1/15$. For a four dimensional Ising lattice ($d$ = 4) on
the other hand, the critical exponents are the mean field exponents\cite{Yeomans}, given
also by fractional values, $\beta_{4D} = 1/2$ and $\delta_{4D}^{-1} = 1/3$.

It is a legitimate question to ask whether for a three-dimensional
Ising lattice ($d$ = 3), for which no general theoretical solution is
available for the moment, the values for $\beta_{3D}$ and $\delta_{3D}^{-1}$
are rational fractions or not. In fact, almost forty years ago, Cyril Domb,
one of the very pioneers in the then rapidly growing field of phase
transitions, suggested that for three-dimensional Ising lattices the
susceptibility critical exponent $\gamma = \beta(\delta - 1)$ might be given by
the fractional value $\gamma_{3D}$ = 5/4 = 1.25. Since then a tremendous amount
of work (experimental, theoretical and computational) has been performed with
the aim to get ever more precise numerical characterizations of the phase
transitions. Table I gives a representative sample\cite{Pelissetto} of
numerical values\cite{Butera,Butera2,Guttmann,Oitman,Blote,Ito,Jasch,Berges}
for the exponents $\gamma_{3D}$ and $\beta_{3D}$ obtained by various methods:
high temperature expansion series, low temperature expansion series, Monte
Carlo simulations and field theoretical methods. The overall picture of
the numerical values for $\gamma_{3D}$ and $\beta_{3D}$ is reasonably good, and
they seem to favor $\gamma_{3D} < 1.25$ and $\beta_{3D} > 0.3125$, but, clearly, the
uncertainties quoted in parentesis cannot be taken strictly at face value.

In
the present work we present results on critical exponents values based upon optimized
accurate Monte Carlo calculations and we investigate to what extent the $d$ = 3
Ising exponents are compatible with the simple fractions interpolated between
the fractional $d$ = 2 Ising exponents and the, fractional too, $d$ = 4 Ising
exponents. In particular we will use direct determinations of $\beta_{3D}$, 
from $M_{s}(T)$ data at $H = 0$, and of $\delta_{3D}^{-1}$, from $M_{c}(H)$
data at $T = T_c$, as well as $\chi^{-1}(T)$ vs $T$, including data both below
$T_c$ (LT phase) and above $T_c$ (HT phase), which allow us to make internal
consistency checks, so that the actual value used for the critical temperature
can be confirmed to be compatible with the scaling requirement
$\gamma' = \gamma$ or not.

In order to establish the reliability of the
data and the propriety of the method of analysis used we will proceed in two
steps. First we will check data on large two-dimensional lattices, for which  
the fractional values of the exponents are known exactly, and then we will
analyze data on large three-dimensional lattices for which the fractional
values proposed are only educated guesses. Our data, therefore, can either
lend support or leave unsupported the fractional values proposed. Figure 1  
shows the evolution of critical exponents with lattice dimensionality 
$d$ = 2, 3, 4,$\dots$ for Ising systems.

%
%
\begin{table*}
\caption{\label{tab:table1} Some estimated $s = 1/2$ critical exponents
for $d$ = 3 Ising simple cubic lattices}
\begin{ruledtabular}
\begin{tabular}{lcccc}
 Method & Reference & Year & $\gamma$ & $\beta$ \\
\hline

 High $T$ expansions & \cite{Butera} & (2002) & 1.2368(10) & 0.3243(30)\footnotemark[1] \\

                     & \cite{Butera2} & (1997) & 1.2388(10) & 0.3278(13)\footnotemark[1]\\

\hline

 Low $T$ expansions & \cite{Guttmann} & (1993) & 1.251(28) & 0.329(9) \\

                     & \cite{Oitman} & (1991) & 1.255(10) & 0.320(3) \\

\hline

 Monte Carlo & \cite{Blote} & (1999) & 1.2372(13)\footnotemark[2] & 0.3269(5) \\

             & \cite{Ito} & (2000) & 1.255(18)\footnotemark[2] & 0.325(5) \\

\hline

 Field Theoretical & \cite{Jasch} & (2001) & 1.2403(8) & 0.3257(5) \\

                   & \cite{Berges} & (1994) & 1.258 & 0.336 \\

\end{tabular}
\end{ruledtabular}
\footnotetext[1]{Obtained using the scaling relation $\beta = \nu(1+\eta)/2$}
\footnotetext[2]{Obtained using the scaling relation $\gamma = (2-\eta)/\nu$}
\end{table*} 
%
%

%
\begin{figure} 
\includegraphics[width=6.1cm,height=7.9cm,angle=270]{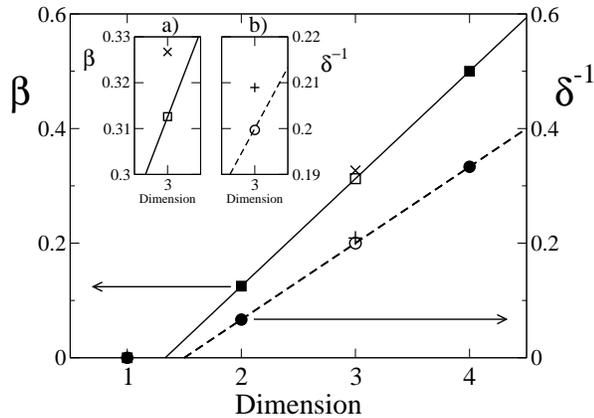}
\caption{\label{fig:epsart1}
 $\beta_{3D}$ and $\delta_{3D}^{-1}$ (open symbols) obtained by 
interpolation between exact two-dimensional 
data from Onsager\cite{Onsager} and four-dimensional data from mean field\cite{Gonzalo}
theory (filled symbols). The crosses show current accepted
results\cite{Butera}. In all cases, error bars are smaller than the
symbol's size. Insets a) and b) are a blow-up close to the $d$ = 3 region
for $\beta$ and $\delta^{-1}$, respectively.}
\end{figure} 
%
%

\section{\label{sec:level1}Monte Carlo determinations of $M_{s}(T)$
and $M_{c}(T)$\protect}

Finite size scaling Monte Carlo simulations of phase transitions are known to be among 
the best and more effective techniques available to characterize the critical
behavior of model systems such as Ising systems of any dimensionality. To
determine critical exponents with the highest possible   reliability it is
desirable (i) to increase systematically the size $L$ of the system till the
results become practically independent of $L$ within error bars; (ii) to make
the interval (temperature or magnetic field) between successive states in
the vicinity of the critical point as small as possible, choosing wisely the
full range (temperature of magnetic field) so that is not too large (time
consuming) or too small (inconvenient of incomplete) to determine adequately
the exponent in question; (iii) to insure that the time spent in the
calculation at each point (temperature, magnetic field) is sufficient to  
arrive to true equilibrium. We did use two-dimensional lattices of $800^2$,
$900^2$ and $1000^2$ spins, and three-dimensional lattices of $90^3$, $100^3$
and $115^3$ spins. The temperature intervals were of the order of 0.001, in
units of $T$ such that $T_{c2D}$ = 2.269185314213... and $T_{c3D}$ =
4.51152(12) for determinations of the spontaneous magnetization $M_{s}(T)$,
and field intervals of the order of 0.0005 for determinations of $M_{c}(H)$ at
the critical isotherm.

The number
of Monte Carlo steps taken to insure equilibrium at each state was 50000 and
the number of states considered in the partition function was 10000. No 
improvement was detected by increasing the number of states. To get
$M_{s}(T)$ a Wolff\cite{Wolff} algorithm was used, which is a generalization of the
Swendsen-Wang\cite{Swendsen} algorithm. To obtain $M_{c}(H)$, a somewhat novel but straightforward,
calculation not usually found in the literature, a standard Metropolis\cite{Metropolis}
algorithm was used and it was proved to be quite adequate.

Initially both periodic boundary
conditions and free boundary conditions were used. After confirming that for
very large lattices the difference between results obtained with either set of boundary
conditions was quite negligible, we used subsequently periodic boundary
conditions all the way.

For the temperature scans we always began at $T > T_c$
and spent sufficient time at the beginning to make sure that thermal
equilibrium was attained already far above the transition temperature. Nevertheless the
statistical ups and downs in the residual $M(T)$ data at $T > T_c$, are much
larger than the corresponding fluctuations at $T < T_c$ which become almost
negligible and display a beautiful continuity all the way down in
temperature. The log-log plots of $M$ vs. $\mid$$T - T_c$$\mid$ and $M$ vs. $H$
indicate clearly, first that we have been able to get really close to the
critical point, more so than in other Monte Carlo calculations for which
data in the literature are sufficiently explicit, and, second, that in the full range
displayed for $T$ or $H$ corrections to scaling are invisible, which make the
data especially apt to determine the exponents $\beta_{3D}$ and
$\delta_{3D}^{-1}$.

\section{\label{sec:level1}Monte Carlo results for two-dimensional lattices
\protect}

Figure 2(a) gives a log-log plot of the spontaneous magnetization as a
function  of temperature for a squared Ising lattice with 1000$\times$1000
spins. Finite size effects show up as rounding at
$\mid$$T - T_c$$\mid \longrightarrow 0$ which occurs only at
$\mid$$T - T_c$$\mid \lesssim 0.002$
with  $T_c$ =  2.269185314213... Corrections to scaling should appear at the
other end of the  temperature range examined, but are quite invisible in our
data. The spontaneous magnetization data give directly a value for
$\beta_{2D}$ = 0.1242 $\pm$ 0.0008 $\cong$ 1/8, as expected. Figure 2(b)
presents a log-log plot of the magnetization as a function  of field for the
same squared Ising lattice with 1000$\times$1000 spins at the critical
isotherm.  Finite size effects begin to appear as incipient rounding at
$H \lesssim 0.0001$ but are almost  imperceptible. Corrections to scaling
should appear at the other end of the field range  investigated but are
completely negligible in our data. These critical isotherm data, not
previously investigated in depth, as far as we know, give  directly
$\delta_{2D}^{-1}$ = 0.06656 $\pm$ 0.00022 $\cong$ 1/15, as expected.

These results lend support to the expectation that carefully taken Monte Carlo
data in large enough lattices taken at small enough temperature / field
intervals are accurate enough to investigate whether critical exponents are
given by simple fractions or not, at least in the case of two dimensions.

A straightforward numerical analysis of the fractions
compatible with the experimental results and the uncertainties quoted 
has been made. For
$d$ = 2 it can be seen that $n/m$ with $n<m$ compatible with the
uncertainties, must go to $m$ values very large, $m > 256$ for $\beta_{2D}$
and $m > 360$ for $\delta_{2D}^{-1}$.

%
%
\begin{figure}
\includegraphics[width=6.1cm,height=7.9cm,angle=270]{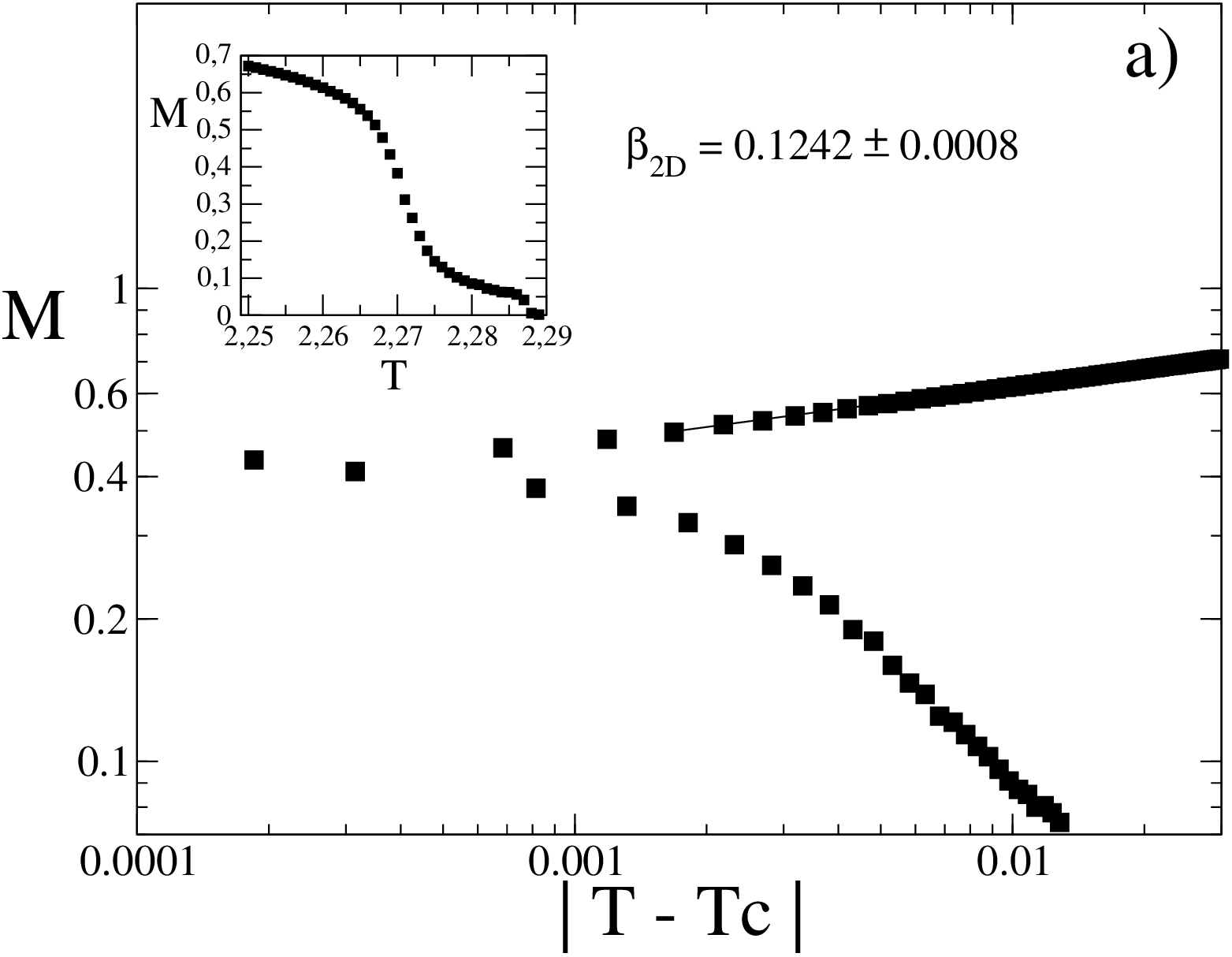}
\\
\includegraphics[width=6.1cm,height=7.9cm,angle=270]{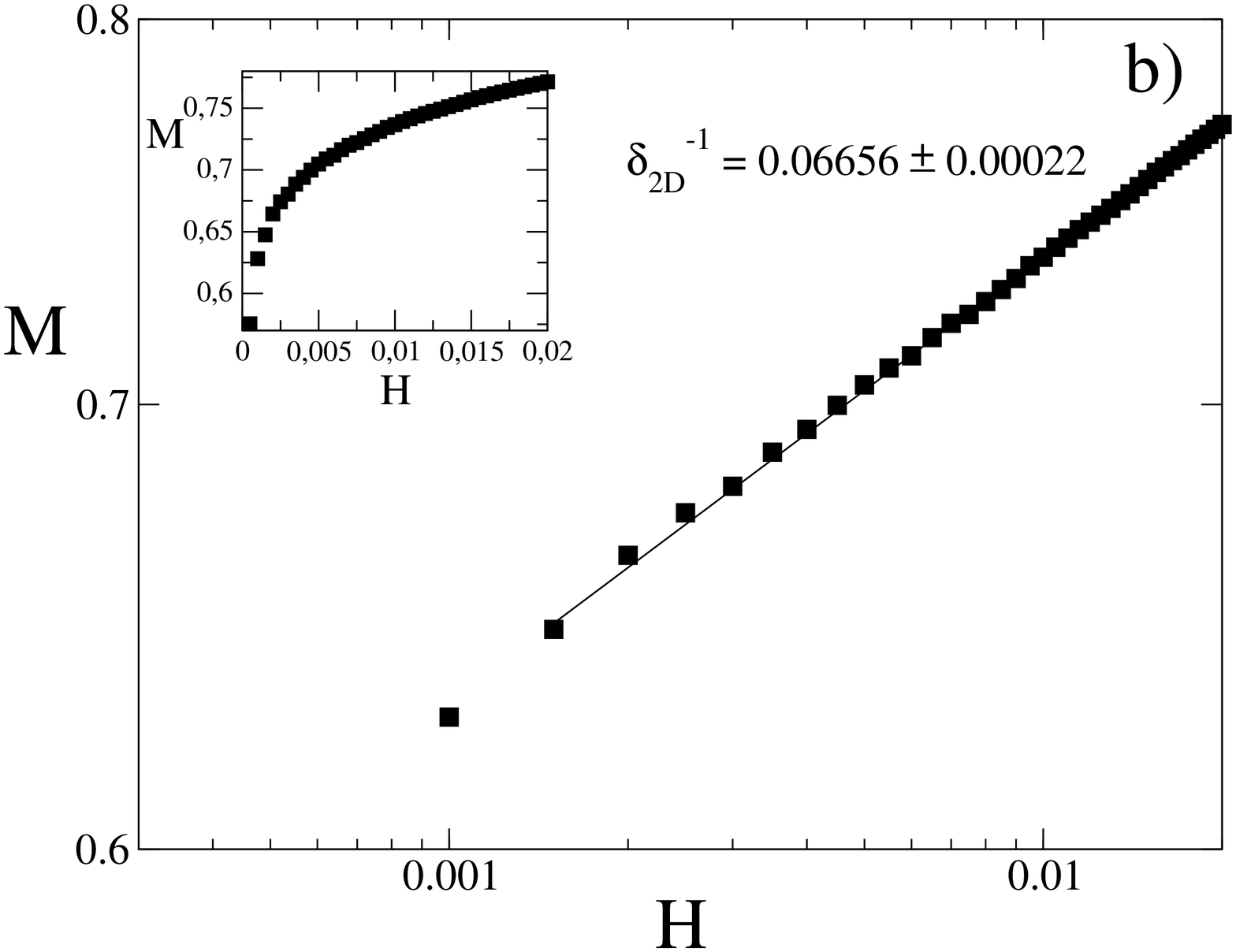}
\\
\includegraphics[width=6.1cm,height=7.9cm,angle=270]{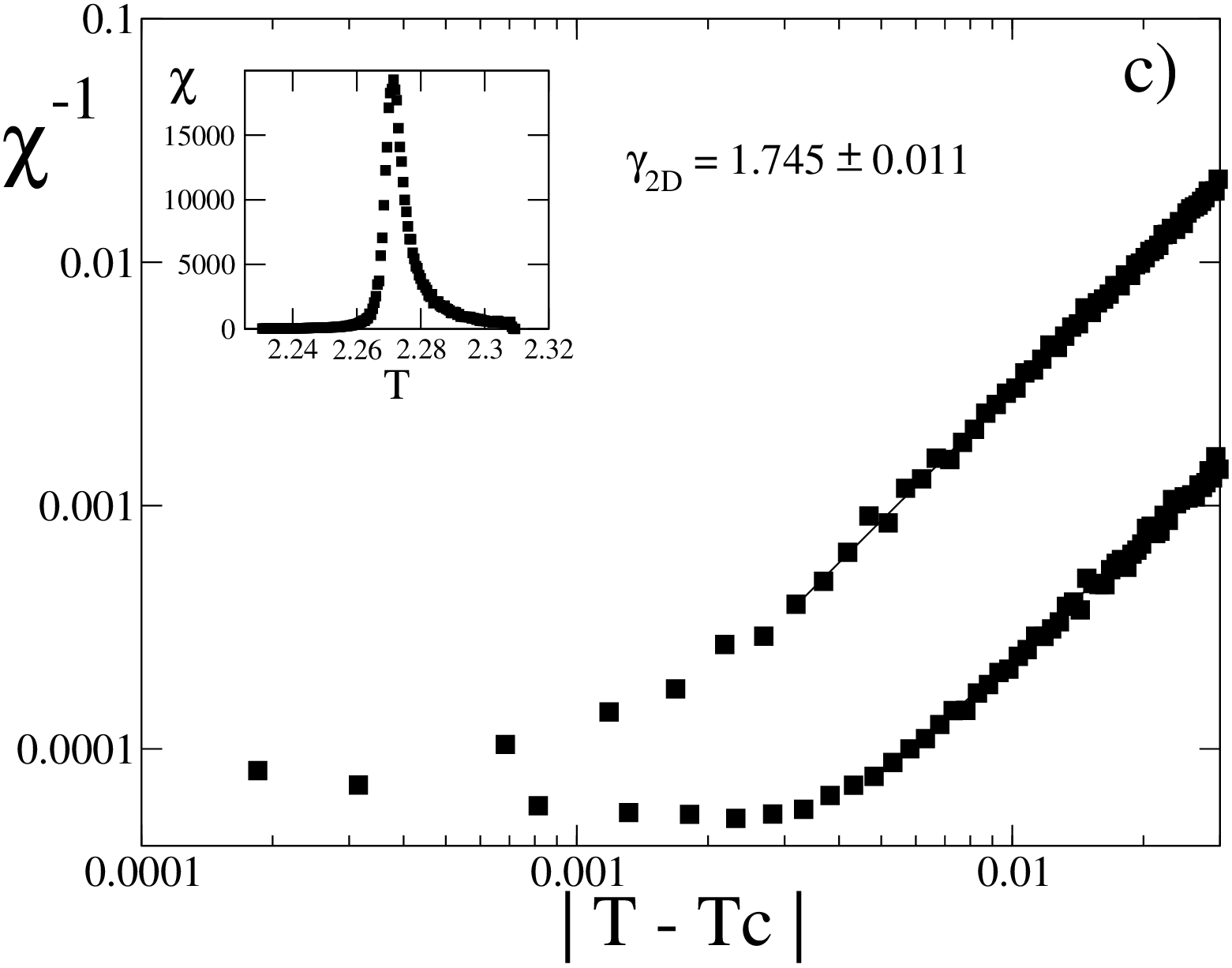}
\caption{\label{fig:epsart} Two dimensional critical
exponents $\beta_{2D}$, $\delta_{2D}^{-1}$ and $\gamma_{2D}$
obtained by fits of Monte Carlo data of spontaneous
magnetizacion $M_{s}(H=0)$ as a function of temperature $T$ near
$T_c = 2.269185314213$, critical isotherm, $M$ as a function of $H$ at
$T = T_{c}$, and isothermal susceptibility $\chi$, respectively,
in a system of $1000\times1000$ spins.
Insets show the row data.}
\end{figure}
%
%

\section{\label{sec:level1}Monte Carlo results for three-dimensional
lattices\protect}

Figure 3(a) gives a log-log plot of $M_{s}(T)$ vs. $T$ for a simple cubic
lattice with 115$\times$115$\times$115 spins. Lattices with
90$\times$90$\times$90 and 100$\times$100$\times$100 spins were investigated
and the results, of course slightly less accurate, where completely
consistent with those obtained with the larger lattice. Again finite size
effects appear at $\mid$$T - T_c$$\mid \longrightarrow 0$,
and they appear clearly at $\mid$$T - T_c$$\mid \lesssim 0.004$ but
corrections to scaling are invisible in the temperature range investigated.
For $\beta_{3D}$ we get

\begin{equation}
\label{eq:beta1}
\beta_{3D} = 0.3126(4) \cong \frac{5}{16}
\end{equation}

Figure 3(b) produces a similar log-log plot of the magnetization as a
function of field for the same lattice with 115$\times$115$\times$115 spins at
$T = T_c$ = 4.51152(12). It is clear that both finite size effects and
systematic departures from scaling are absent or imperceptible in the field
range explored. These new, not previously directly investigated critical
isotherm data, result in

\begin{equation}
\label{eq:delta1}
\delta_{3D}^{-1} = 0.1997(4) \cong \frac{1}{5}
\end{equation}

which differs somewhat from previous indirect estimates of
$\delta_{3D}$ summarized by A. Pelissetto and E. Vicari\cite{Pelissetto} as
giving $\delta_{3D}$ = 4.789(2), equivalent to $\delta_{3D}^{-1}$ = 0.2088(1).

Using the numerical values for $\beta_{3D}$ and $\delta_{3D}^{-1}$ in Equations
(\ref{eq:beta1}) and (\ref{eq:delta1}) one gets $\gamma_{3D}$ indirectly as

\begin{equation}
\label{eq:gamma1}
\gamma_{3D} = \beta_{3D}(\delta_{3D} - 1) = 1.252(4)
\end{equation}

by means of the scaling relation for $\gamma$ in terms of $\beta$ and
$\delta$.

But one can get $\gamma$ directly from plots of susceptibility data at
$T > T_{c} (\gamma)$ as well as at $T < T_{c} (\gamma')$ giving us the
opportunity to check that the fundamental equality $\gamma = \gamma'$,
confirmed by renormalization group theory\cite{Pelissetto}, is fulfilled.

Figure 3(c) gives log-log plots of direct Monte Carlo data on the
susceptibility $\chi^{-1}(T)$ as a function
of $\mid$$T - T_c$$\mid$ both above and below the critical
temperature $T_{c}$ =
4.51152. The resulting directly determined value of gamma is

\begin{equation}
\label{eq:gamma2}
\gamma_{3D} = \gamma_{3D}' = 1.253(4)
\end{equation}

which is in excellent agreement with the forty-year-old prediction of Domb and
Sykes\cite{Domb}.

As mentioned before the following consistency check was
made: changing smoothly the critical temperature value by increments
(decrements) of 0.00002 we can obtain smooth changes in the effective values
of the exponents $\gamma_{3D}$ and $\gamma_{3D}'$ to reproduce numerical values
as low as 1.237 for $\gamma_{3D}$, resulting in $\gamma_{3D}' > 1.25$, or
as high as 1.263 for $\gamma_{3D}$, resulting in $\gamma_{3D}' < 1.25$. With
our Monte Carlo data, which in sufficiently wide ranges appear to be free of
finite size effects as well as of corrections to scaling effects, only using
the right $T_c$ = 4.51152, in perfect agreement with the values quoted in the
literature, the requirement $\gamma = \gamma'$ is duly fulfilled.

%
%
\begin{table}

\caption{\label{tab:table2}
Monte Carlo $s = 1/2$ Ising critical exponents in a three-dimensional
simple cubic lattice of $115\times115\times115$ spins
}
\begin{ruledtabular}
\begin{tabular}{lc}

$\beta$ & $n/m$\footnotemark[3]\\
0.31254$\pm$0.00029 & 5/16, ..., 40/128   (41/131) \\

\hline

$\delta^{-1}$ & $n/m$\footnotemark[3] \\
0.1997$\pm$0.0004 & 1/5, ..., 57/285   (57/286) \\

\end{tabular}
\end{ruledtabular}
\footnotetext[3]{Fractional values compatible with the quoted uncertaintities
up to $m_{max} = 128$ for $\beta$ and up to $m_{max} = 285$ for $\delta^{-1}$.
More complex fractions (shown in parenthesis) are compatible with the 
uncertaintities only with denominators larger than 128 and 285, respectively.}

\end{table}
%
%

Table II shows that
according to our data for simple cubic $s$ = 1/2 Ising lattices, simple
fractions for $\beta$ and $\delta^{-1}$ (the same procedure could be employed to
check simple fractions for $\gamma$) compatible with the quoted uncertainties
are given always by $n/m$ fractions which are, either 
identical to the interpolated values
$\beta_{3D}$ = 5/16, $\delta_{3D}^{-1}$ = 1/5 ($\gamma_{3D}$ = 5/4) or much more
complex fractions involving $m > 128$ (for $\beta$) and $m > 285$ (for
$\delta^{-1}$).

%
%
\begin{figure}
\includegraphics[width=6.1cm,height=7.9cm,angle=270]{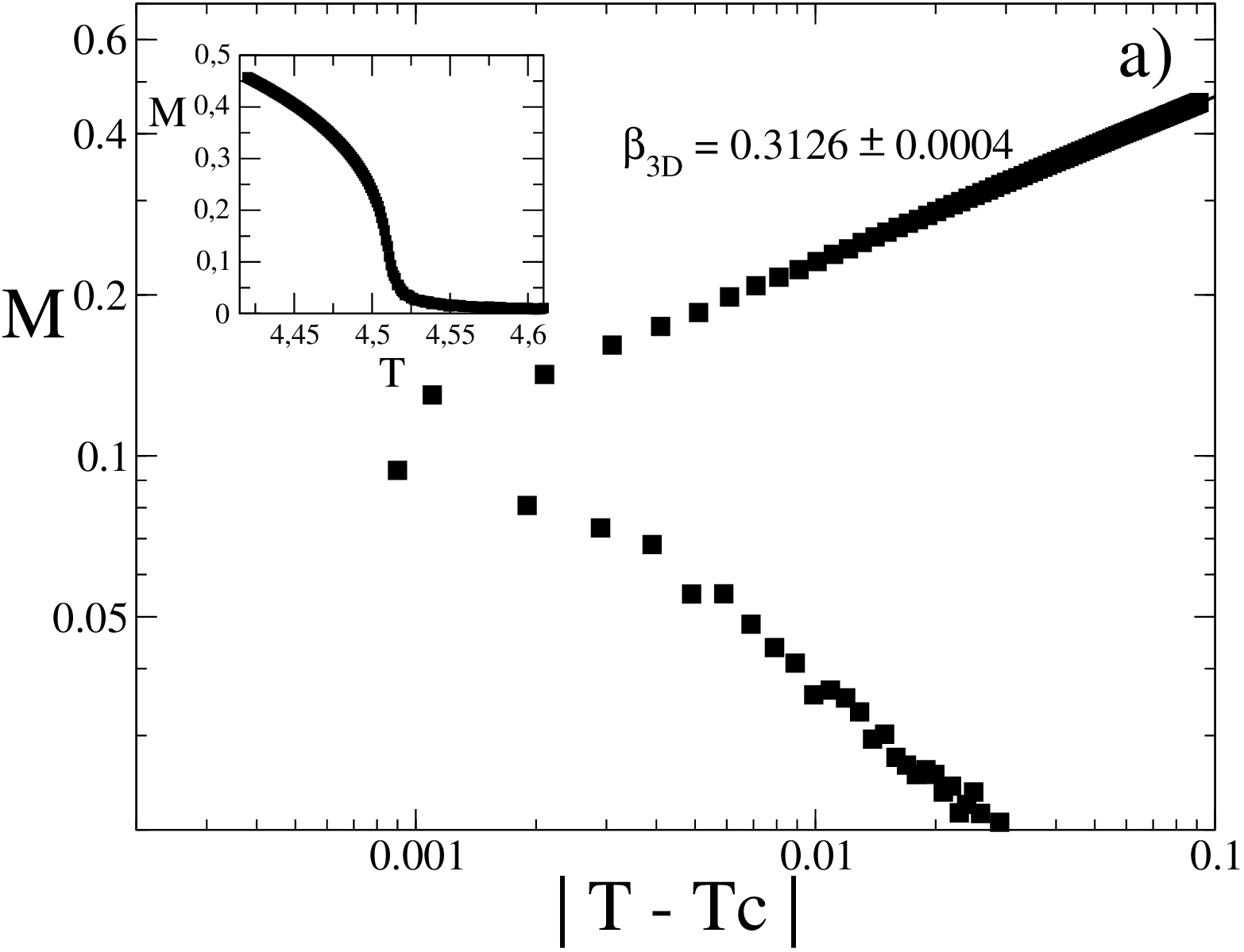}
\\
\includegraphics[width=6.1cm,height=7.9cm,angle=270]{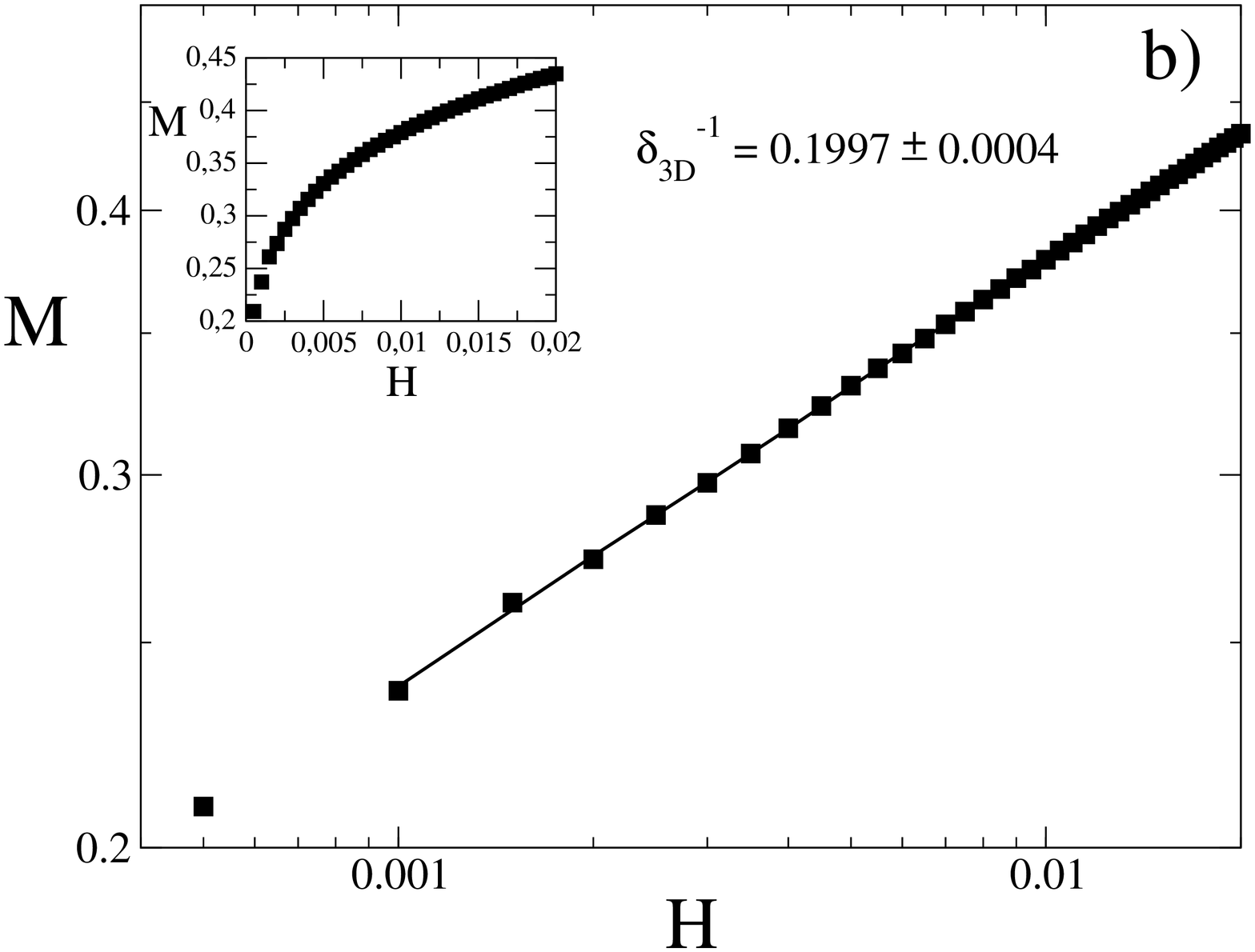}
\\
\includegraphics[width=6.1cm,height=7.9cm,angle=270]{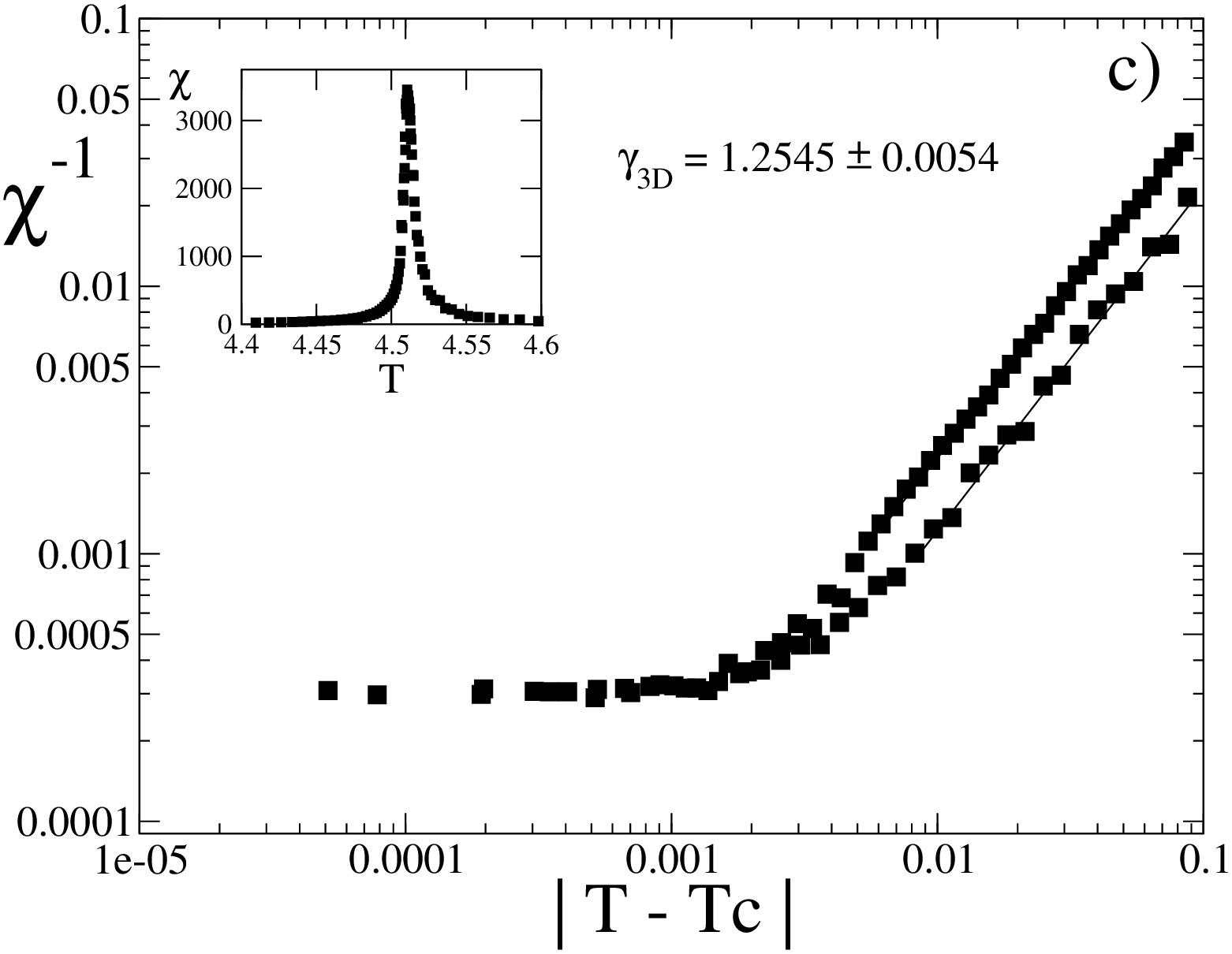}
\\
\caption{\label{fig:epsart3} Three-dimensional critical exponents $\gamma_{3D}$ (a),
$\beta_{3D}$ (b) and $\delta_{3D}^{-1}$ (c) obtained by fits of Monte Carlo data
of spontaneous magnetization $M_{s}$ near $T_c$
= 4.51152 $\pm$ 0.00012, critical isotherm and isothermal susceptibility 
$\chi$, respectively, in a system of
$L^{3}$ = 115$\times$115$\times$115 spins with periodic boundary conditions.
Insets show the row data.}
\end{figure}
%
%

\section{\label{sec:level1}Concluding remarks\protect}
Our Monte Carlo data do not prove directly beyond doubt that
three-dimensional Ising lattices are characterized by the fractional
critical exponents $\beta_{3D}$ = 5/16 and $\delta_{3D}^{-1}$ = 1/5 (and
through the corresponding scaling relations, by the resulting critical
exponents $\gamma = 5/4$, $\alpha = 1/8$, $\nu = 5/8$ and $\eta = 0$)
but they support
strongly and consistently this fractional values, in particular
the susceptibility data resulting
in $\gamma_{3D}$ = $\gamma_{3D}'$ = 1.25 = 5/4.
Our empirical data may be the basis for future well grounded theoretical
arguments confirming the above fractional critical
exponents.

\begin{acknowledgments}
We acknowledge helpful comments by Lidia Braunstein and Gerry Paul to
Jorge Garc\'{\i}a and Manuel I. Marqu\'es during fruitful stays at the
Boston University Physics Department.
Support from the Spanish Ministry of Science and Technology through
Grant Number BFM2000-0032 is gratefully acknowledged.
M.I. Marqu\'es acknowledges the Postdoctoral Grant at Boston
University
 from the Spanish Ministry of Education.
\end{acknowledgments}

\newpage 
\bibliography{apssamp}

\begin{thebibliography}{99}

\bibitem{Adler}
See f.i. J. Adler,
{\em J. Phys. A\/} {\bf 16}, 3585 (1983).

\bibitem{Onsager}
L. Onsager,
{\em Phys. Rev\/} {\bf 65}, 117 (1944).

\bibitem{Yeomans}
J.M. Yeomans,
{\em Statistical Mechanics of Phase Transitions\/},
(Oxford University Press, 1992).

\bibitem{Pelissetto}
See f.i. A. Pelissetto and E. Vicari,
{\em Phys. Rep\/} {\bf 368}, 549-727 (2002).

\bibitem{Butera}
P. Butera and M. Comi,
{\em Phys. Rev. B\/} {\bf 65}, 144431 (2002).
\bibitem{Butera2}
P. Butera and M. Comi,
{\em Phys. Rev. B\/} {\bf 56}, 8212 (1997).

\bibitem{Guttmann}
A.J. Guttmann and I.G. Enting,
{\em J. Phys. A} {\bf 26}, 806 (1993).

\bibitem{Oitman}
J. Oitman, C.J. Hamer and W. Zheng,
{\em J. Phys. A} {\bf 24}, 2863 (1991).

\bibitem{Blote}
H.W.J. Bl\"ote, L.N. Shchur and A.L. Talapov,
{\em Int. J. Mod. Phys. C\/} {\bf 10}, 137 (1999).

\bibitem{Ito}
N. Ito, K. Hukushima, K. Ogawa and Y. Ozeki,
{\em J. Phys. Soc. Japan\/} {\bf 69}, 1931 (2000).

\bibitem{Jasch}
F. Jasch and H. Kleinert,
{\em J. Math. Phys.\/} {\bf 42}, 52 (2001).

\bibitem{Berges}
J. Berges, N. Tetradis and C. Wetterich,
{\em Phys. Rev. Lett.\/} {\bf 77}, 873 (1996).

\bibitem{Caracciolo}
S. Caracciolo, G. Ferraro and A. Pelissetto,
{\em J. Phys. A\/} {\bf 24}, 3625 (1991).

\bibitem{Wolff}
U. Wolff,
{\em Phys. Rev. Lett.\/} {\bf 62}, 361 (1989).

\bibitem{Swendsen}
R.H. Swendsen and J.S. Wang,
{\em Phys. Rev. Lett.\/}. {\bf 58}, 86 (1987).

\bibitem{Metropolis}
N. Metropolis, A.W. Rosenbluth, M.N. Rosenbluth, A.H. Teller and E. Teller,
{\em J. Chem. Phys.\/} {\bf 21}, 1087 (1953).

\bibitem{Domb}
Domb and Sykes, 
{\em Phys. Rev.\/} {\bf 128}, 168-173 (1962).



\bibitem{Domb2}
C. Domb, in 
{\em Phase Transitions and Critical Phenomena\/},
edited by C. Domb and M.S. Green (Academic, London, 1974).

\bibitem{Domb3}
C. Domb, {\em The Critical Point\/} (Taylor\&Francis, London, 1996).

\bibitem{Fisher2}
M.E. Fisher, {\em Rev. Mod. Phys.\/} {\bf 46}, 597 (1962); {\bf 70}, 653 (1998).

\bibitem{Gonzalo}
See f.i. J.A. Gonzalo,
{\em Effective Field Approach to Phase Transitions and some
Applications to Ferroelectrics\/}
(World Scientific, Singapore, 1991).



\end{thebibliography}

\end{document}